\documentclass[10pt,journal,compsoc]{IEEEtran}
\ifCLASSOPTIONcompsoc
  \usepackage[nocompress]{cite}
\else
  \usepackage{cite}
\fi
\ifCLASSINFOpdf
   \usepackage[pdftex]{graphicx}
\else
\fi
\usepackage{dsfont}
\usepackage{amsmath,amssymb}
\usepackage{lipsum}

\DeclareMathOperator*{\argmax}{\arg\!\max}

\usepackage{algorithm}
\usepackage{varwidth}
\usepackage{algpseudocode}

\makeatletter
\newcommand{\StatexIndent}[1][3]{%
	\setlength\@tempdima{\algorithmicindent}%
	\Statex\hskip\dimexpr#1\@tempdima\relax}
\algdef{S}[WHILE]{WhileNoDo}[1]{\algorithmicwhile\ #1}%
\makeatother

\algnewcommand\algorithmicinput{\textbf{Input:}}
\algnewcommand\algorithmicoutput{\textbf{Output:}}
\algnewcommand\algorithmicbl{\textbf{\# of blocks:}}
\algnewcommand\algorithmicth{\textbf{\# of threads / block:}}
\algnewcommand\Input{\item[\algorithmicinput]}
\algnewcommand\Output{\item[\algorithmicoutput]}
\algnewcommand\Block{\item[\algorithmicbl]}
\algnewcommand\Thread{\item[\algorithmicth]}
\algrenewcommand\algorithmicindent{1.0em}

\usepackage{pifont}
\usepackage{tabularx}
\usepackage{array,booktabs}
\usepackage{multirow}
\usepackage{slashbox}
\usepackage{dblfloatfix}
\newcolumntype{C}{>{\centering\arraybackslash}X} 
\usepackage{lipsum}

\usepackage{soul}
\usepackage{array}
\usepackage{color}
\usepackage{nicefrac}

\newcolumntype{P}[1]{>{\centering\arraybackslash}p{#1}}

\newcommand{\sslash}{\mathbin{/\mkern-2mu/}}

\newcommand{\mRefAlg}[1]{Algorithm~\ref{#1}}
\newcommand{\mRefFig}[1]{Fig.~\ref{#1}}
\newcommand{\mRefEq}[1]{Equation~\ref{#1}} 

\newcommand{\mRefSec}[1]{Section~\ref{#1}}

\usepackage{hyperref}
\begin{document}
%
\title{gIM: GPU Accelerated RIS-based Influence Maximization Algorithm}
%
%
%

\author{Soheil~Shahrouz,~
        Saber~Salehkaleybar,~
        and~Matin~Hashemi
        
        	{\color{blue} 
        	\begin{flushleft}
        		\footnotesize 
        		This article is published. Please cite as S. Shahrouz, S. Salehkaleybar, M. Hashemi,  ``gIM: GPU Accelerated RIS-based Influence Maximization Algorithm," IEEE Transactions on Parallel and Distributed Systems.
        \end{flushleft} }  
        
\thanks{
Authors are with Learning and Intelligent Systems Laboratory, Department
of Electrical Engineering, Sharif University of Technology, Tehran, Iran. \protect \\ 
E-mails: soheil.shahrouz@ee.sharif.edu, saleh@sharif.edu, matin@sharif.edu.  Webpage: http://lis.ee.sharif.edu 
}
\thanks{(Corresponding author: Saber~Salehkaleybar)}
}

\IEEEtitleabstractindextext{%
\begin{abstract}
Given a social network modeled as a weighted graph $G$, the influence maximization problem seeks $k$ vertices to become initially influenced, to maximize the expected number of influenced nodes under a particular diffusion model. The influence maximization problem has been proven to be NP-hard, and most proposed solutions to the problem are approximate greedy algorithms, which can guarantee a tunable approximation ratio for their results with respect to the optimal solution. The state-of-the-art algorithms are based on Reverse Influence Sampling (RIS) technique, which can offer both computational efficiency and non-trivial $(1-\nicefrac{1}{e}-\epsilon)$-approximation ratio guarantee for any $\epsilon >0$. RIS-based algorithms, despite their lower computational cost compared to other methods, still require long running times to solve the problem in large-scale graphs with low values of $\epsilon$. 
In this paper, we present a novel and efficient parallel implementation of a RIS-based algorithm, namely IMM, on GPU. The proposed GPU-accelerated influence maximization algorithm, named gIM, can significantly reduce the running time on large-scale graphs with low values of $\epsilon$.
Furthermore, we show that gIM algorithm can solve other variations of the IM problem, only by applying minor modifications. Experimental results show that the proposed solution reduces the runtime by a factor up to $220 \times$. 
The source code of gIM is publicly available online.
\end{abstract}

}

\maketitle

\IEEEdisplaynontitleabstractindextext

%
\IEEEpeerreviewmaketitle

\section{Introduction}
\label{sec:intro}
\IEEEPARstart{W}{ith} the explosion of social network services in the last decades, hundreds of millions of people can easily interact with each other. The vast prevalence of social networks facilitates large-scale viral marketing through "word-of-mouth" effects, where each individual recommends a product to his/her friends, contributing to a large number of adoptions of the product. In order to succeed in a viral marketing campaign, it is required to choose a few \textit{influential} individuals and provide incentives (e.g. free samples of the product and/or cash) to create a cascade of product adoptions as large as possible. Obtaining such a set of users is commonly known as the ``influence maximization (IM) problem". In particular, for a given network $G$ and a probability model representing diffusion mechanism in the network, IM aims to select a set of users (called seed set) which maximizes the expected number of users that are affected in the diffusion process. 
 
In a seminal work, Kempe et al. \cite{kempe2003maximizing} formulated the IM problem as a combinatorial optimization problem for two popular diffusion models, namely, independent cascade (IC) and linear threshold (LT). They showed that the IM problem is NP-hard in both models, and proposed a greedy framework that can obtain $(1-\frac{1}{e}-\epsilon)$-approximate solutions for any $\epsilon>0$. More specifically, they considered the influence spread (the expected number of affected users) as a function of seed set. Starting from an empty set, a user with the maximum marginal gain to the influence spread is added to the seed set in an iterative manner until the desired number of users are selected. Almost all the proposed algorithms for the IM problem followed this greedy framework. Later, it has been shown that computing the influence spread as a function of seed set is $\#$P-hard under both IC and LT models \cite{chen2010scalablelt, chen2010scalableic}.  These theoretical results have motivated many researchers to design efficient estimators for the influence spread. In this regard, existing algorithms for the IM problem can be classified into three main approaches: I) simulation-based approach, II) proxy-based approach, and III)  sketch-based approach.

In the simulation-based approach, Monte Carlo (MC) simulations are executed to estimate the influence spread for a given seed set by averaging the number of affected users over all the generated instances. This approach was considered in \cite{kempe2003maximizing} for the first time and subsequent works tried to reduce the number of MC simulations through \textit{lazy evaluation} \cite{leskovec2007cost, goyal2011celf++}. The main advantage of this approach is that it can be utilized for any diffusion model. However, it is not scalable to very large graphs since it requires to generate too many instances to compute the influence spread with a desirable estimation error. 
In the proxy-based approach, the main idea is to use simple models, such as shortest path or PageRank, instead of complicated diffusion models. This reduces the runtime for computing the influence spread significantly. However, improving the time complexity of evaluating the influence spread comes at the expense of losing theoretical guarantees on estimation error bounds. In fact, it has been shown that the proxy-based approach might have unstable behavior in some  graphs in the sense that the influence spread could change significantly by a small modification in the graph structure \cite{he2014stability}. 
The sketch-based approach has been proposed to overcome computational inefficiency issues of the simulation-based approach and instability problems of the proxy-based approach. The main idea is to first generate \textit{sketches} for a given diffusion model. Afterwards, the influence spread is estimated based on the generated sketches. Due to the desired properties of this approach, many recent algorithms for the IM problem have focused on efficiently generating sketches (see \mRefSec{sec:prelim}). 

Although the most recent algorithms for the IM problem are mainly based on constructing sketches, they still have computational issues for small estimation error bounds in large graphs. Moreover, this problem might become critical in some recent applications of IM such as multi-round IM \cite{sun2018multi}, where it is required to solve the IM problem multiple times for a given graph. 
 
The aim of this paper is to accelerate the state-of-the-art sketch-based algorithm (IMM \cite{tang2015influence}) using parallel processing on GPU. Hence, the proposed solution is named gIM. Our main contributions are as follows:
\begin{itemize}
	\item One of the key parts in IMM algorithm is to find a reverse reachable set of nodes from a randomly selected node by executing a BFS algorithm. This part of the algorithm is repeated multiple times. Many common approaches for the acceleration of BFS on GPU run only one BFS at a time, and therefore, produce a single reverse reachable set. 
	Here, we propose a method to simultaneously generate multiple reverse reachable sets, and at the same time, parallelize each task of generating a reverse reachable set as well. More specifically, several BFSs are executed in parallel, and in each one, adjacent edges of a node are processed in parallel, but nodes in the frontier queue are processed sequentially (Section \ref{sec:alg:rr}). 
	\item We also propose methods to judiciously store large frontier queues of the BFSs in limited GPU shared memory in order to further reduce the runtime (Section \ref{sec:alg:rr:overflow}).
	\item We propose data structures and methods to optimally store the sketches in GPU global memory (Section \ref{sec:alg:rr:n_blocks}) and to process them efficiently in order to select the seed set (Section \ref{sec:alg:max_cover}). This helps to fit the required internal data structures into GPU global memory for large graphs with millions of nodes.
	\item We discuss how to optimize the required synchronization mechanism and also the number of threads and blocks in order to exploit all processing power of GPU (Section \ref{sec:alg:opt}).
	\item The proposed parallelization of IMM algorithm on GPU can be utilized in other applications that use a variant of IMM as a subroutine. For instance, we show that our implementation can be adapted to the multi-round IM problem. The experiments show that running times are greatly reduced (Section \ref{sec:exp:MR}).
	\item Experimental results on real social networks show that the proposed gIM algorithm can outperform state-of-the-art solutions by a factor of up to $220\times$.
\end{itemize}

	In the following, we briefly review some of the most prominent works in the literature, which focused on parallel implementation of the IM algorithm. 
	%
	Liu et al. \cite{liu2013imgpu} presented the IMGPU framework to accelerate the basic greedy algorithm with MC simulations on GPU. They first converted each instance graph to a DAG to avoid redundant graph traversals in the computations of each node's influence spread. Next, they designed a bottom-up traversal algorithm to compute the marginal gain of all nodes in the resulting DAG on GPU hardware. IMGPU reduces the execution time of the greedy algorithm by a factor of $60$x but it still needs tens of minutes to solve the problem in relatively large graphs. 
	
	Pal et al. \cite{pal2017fast} focused on GPU acceleration of the greedy algorithm with MC simulations under continuous-time diffusion models. They assumed that each node's influence spread is limited to its local neighborhood and restricted the graph traversal to this small neighborhood. They adopted a node-level parallelization approach for calculating marginal gains. Despite achieving outstanding speedups, the limited graph traversal degrades the seed set quality, and hence, this method does not reach the same influence spread as the state-of-the-art algorithms.
	
	Minutoli et al. \cite{minutoli2019fast} proposed Ripples, a framework for multi-core and distributed implementation of the IMM algorithms. Later, by introducing cuRipples \cite{minutoli2020curipples} framework,  they extended their work to provide support for multi-GPU systems and employed GPUs along with multi-core CPUs to further reduce the execution time. 
	
	Gokturk et al. \cite{gokturk2020boosting} accelerated simulation-based algorithms on multi-core CPUs by employing a direction-oblivious hash-based sampling approach to fuse instance graph generation and marginal gain calculation. They also exploited SIMD vectorization to run multiple concurrent simulations on a single core. However, direction-oblivious sampling causes their algorithm to be applicable only on undirected graphs. 
	
	Nguyen et al. \cite{nguyen2019blocking} proposed a CPU-GPU method to accelerate Spread Interdiction (SI) problem which is similar to the IM problem. Although they used concepts similar to reverse influence sampling, their proposed method is not directly applicable to the IM problem.
	
	In a parallel line of research, there have been efforts to make large-scale graph processing more tractable by graph sampling, that is, producing a graph of much smaller size while preserving the desirable characteristics of the original graph. Pandey et al. \cite{pandey2020c} proposed C-SAW framework to implement various graph sampling methods and paralleled them on GPU. Zeng et al. \cite{zeng2019graphsaint} introduced GraphSAINT, a graph sampling method to train graph convolutional networks (GCN), which paves the way for training GCNs on large graphs and thus opens the opportunity to use GCN for solving the IM problem.

The rest of this paper is organized as follows: \mRefSec{sec:prelim} reviews the preliminaries. \mRefSec{sec:alg} presents our proposed GPU-accelerated algorithm for the IM problem, named gIM. \mRefSec{sec:exp} presents experiment results and comparisons with related works.  \mRefSec{sec:conc} concludes the paper.

\section{Preliminaries}
\label{sec:prelim}
In this section, we introduce some notations and investigate two common diffusion models, namely independent cascade and linear threshold models. Then we formally define the influence maximization problem and review Kempe's greedy algorithm \cite{kempe2003maximizing} and methods based on reverse influence sampling.

\subsection{Notation}
\label{sec:prelim:not}

Directed graph $G=(V,E)$ models a social network, where $V$ denotes the set of vertices (i.e., users) and $E$ indicates the set of directed edges (i.e., relationships between users). For every edge $e=(u,v)\in E$, we define $u$ as an incoming neighbor of $v$, and $v$ as an outgoing neighbor of $u$. Also, we denote sets of \textbf{incoming} and \textbf{outgoing} neighbors of every node $u$, by $N_I(u)$ and $N_O(u)$, respectively. Moreover, we assume that every edge $e=(u,v)\in E$ is associated with an \textbf{influence probability}  $p_{uv}\in[0,1]$. We denote the total number of nodes and edges in $G$ by $n$ and $m$, respectively.

\subsection{Diffusion Models}
\label{sec:prelim:dm}

There is an immense amount of literature on models that capture the diffusion phenomenon's behavior \cite{wen2014sword, chen2012time, kim2014ct, xie2015dynadiffuse}. Among these models, independent cascade (IC) and linear threshold (LT) models have been extensively studied. In the following, we briefly discuss these models. 

\subsubsection*{Independent Cascade (IC) Diffusion Model}
\label{sec:prelim:ic}

Given a graph $G=(V,E)$, edge influence probabilities $p_{uv}$ for every edge $e=(u,v)$, and a subset of vertices $S\subseteq V$, which is commonly called \textbf{seed set}, an instance of influence diffusion process under the IC model is generated as follows:

The influence propagates in discrete time steps. At time $t_0$, only the nodes in the seed set $S$ are activated, and all the other nodes remain inactive. At every time step $t_{i+1}$, every node $u$, which has been activated at time step $t_i$, has a chance to activate its inactive outgoing neighbors. Node $u$ succeeds to activate an inactive outgoing neighbor $v$, with probability $p_{uv}$. When a node is activated (either at $t_0$ or by an incoming active neighbor), it remains active in the following time steps. The influence diffusion process terminates at a time step at which none of the previously activated nodes can activate a new inactive node.

\subsubsection*{Linear Threshold (LT) Diffusion Model}
\label{sec:prelim:lt}

The LT model imposes a new restriction on edge influence probabilities, by which the sum of probabilities of incoming edges for every node $u$, must be less than or equal to $1$. Moreover, in the LT model, every node $u$ is associated with a threshold $\tau_u$, which is a real value in the range  $[0,1]$. Given a seed set $S$, a single diffusion process instance under LT model can be described as follows:

Similar to the IC model, the influence propagates in discrete time steps. At time step $t_0$, we assign a random number sampled uniformly from $[0,1]$ to each node $u$, as its threshold $\tau_u$. Furthermore, we initially activate every node $u\in S$ and leave the other nodes as inactive. At time step $t_{i+1}$, every node that has been previously activated remains active, and every inactive node $v$ becomes active when it satisfies the following condition
\begin{equation}
\sum_{u \in N_I(v)} p_{uv}.\mathds{1}_A(u) \geq \tau_v,
\label{eq:lt_act_cond}
\end{equation}
where $A$ is the set of all activated nodes until some time step $t_i$, and $\mathds{1}_A(u)$ is indicator function which is equal to one if $u\in A$. The influence diffusion process stops at a time step at which no inactive node can become active.

\subsection{Problem Statement}
\label{sec:prelim:problem}

The \textbf{influence spread} of seed set $S$ in a single influence diffusion process instance is defined as the total number of active nodes after the termination of the diffusion process, and is denoted by $I(S)$. It should be noted that $I(S)$ is not a deterministic function of $S$. The influence diffusion procedure is a random process which depends on edge influence probabilities (and also node thresholds in LT model). Therefore, $I(S)$ is a random variable.

Given graph $G$, edge influence probabilities $p_{uv}$, a particular diffusion model, and a constant positive integer $k$, the goal of the influence maximization problem is to find seed set $S^*\subset V$ of size $k$, which has the maximum \textbf{expected influence spread} $ \mathbf{E} [ I(S) ]$. The problem can also be expressed as follows:
\begin{equation}
S^* = \argmax_{S\subset V} \left\{ \mathbf{E} [ I(S) ] \mid |S| = k \right\},
\end{equation}
where $S^*$ is the optimal seed set with size $k$. 
It has been shown \cite{kempe2003maximizing} that the IM problem is NP-Hard under both IC and LT diffusion models. 

\subsection{Basic Greedy Algorithm}
\label{sec:prelim:gre}

Kempe et al. \cite{kempe2003maximizing} proposed a greedy algorithm to solve the IM problem. This algorithm  iteratively selects nodes for insertion into the seed set. First, the seed set $S$ is initialized by an empty set. In every iteration, the node $v$ whose insertion into the seed set $S$ leads to the largest increase in $ \mathbf{E} [ I(S) ]$ is selected. The amount of increase is called \textbf{marginal gain}. 
The algorithm terminates after $k$ iterations, i.e., when the size of seed set $S$ reaches $k$.

The greedy algorithm has a simple structure, but its main drawback is that it requires $ \mathbf{E} [ I(S \cup u) ]$ to be computed in every iteration for all the remaining nodes. Unfortunately, evaluating $ \mathbf{E} [ I(S) ]$ has been proven to be \#P-Hard \cite{chen2010scalablelt}. To tackle this problem, Kempe et al. \cite{kempe2003maximizing} proposed a method based on Monte Carlo (MC) simulations in order to estimate $\mathbf{E} [ I(S \cup u) ]$. In order to execute a single MC simulation under IC diffusion model, an \textbf{instance graph} $g$ is created from $G$, by removing every edge $(u,v)\in E$ by probability $1-p_{uv}$. \textbf{Reachable set} $R(S)$ is defined as the set of all nodes that are reachable from $S$ in the resulted graph $g$. Kempe et al. have shown that $\mathbf{E}[| R(S)|] = \mathbf{E}[I(S)]$, and therefore, one can estimate $\mathbf{E}[I(S)]$ by determining $\mathbf{E}[|R(S)|]$ through running a large number of MC simulations and generating many instances of $g$. 
They also proved that by knowing the exact values of $\mathbf{E}[I(S)]$, greedy algorithm’s solution has an approximation ratio of $(1-\frac{1}{e})$. However, in practice, the value of $\mathbf{E}[I(S)]$ is estimated and the true approximation ratio is $(1 - \frac{1}{e} - \epsilon)$, in which $\epsilon$ depends on both graph $G$ and the total number of MC simulations. 
In \cite{kempe2003maximizing}, they did not provide any formal analysis on the number of required MC simulations in order to achieve $(1- \frac{1}{e} - \epsilon)$ approximation ratio and only suggested using $10,000$ MC simulations. Several following works also used $10,000$ as the number of MC simulations without providing any formal analysis. Later, Chen et al. \cite{chen2013information} conducted an analysis on the total number of MC simulations and relative error $\epsilon$. They showed that to achieve $(1- \frac{1}{e} - \epsilon)$ approximation ratio under both IC and LT diffusion models, with a probability of $1-\frac{1}{n}$, one needs to execute $\Theta(\epsilon^{-2}k^2 n\log(n^2k))$ number of MC simulations. 
Therefore, lower values of $\epsilon$ requires larger number of MC simulations.

\subsection{RIS-based Algorithms}
\label{sec:prelim:ris}

As stated above, in every iteration, the greedy algorithm evaluates the marginal gain (i.e., the improvement in expected influence spread) resulting from adding one node which is not currently in the seed set. Thus, the number of marginal gain evaluations is $O(n)$, and because there are $k$ iterations in total, the greedy algorithm conducts $O(nk)$ marginal gain evaluations. Every marginal gain evaluation involves calculating $\mathbf{E}[|R(S)|]$, which is a computationally expensive task. In addition, in every iteration, we are only interested in the node with the largest marginal gain, therefore, marginal gain evaluations for all the other nodes are virtually wasted and are not useful in subsequent iterations.

To address this problem, Borgs et al. \cite{borgs2014maximizing} proposed a method called \textbf{Reverse Influence Sampling (RIS)}, by which there is no need to run many MC simulations for every remaining node in every iteration in order to evaluate the marginal gain. 

In order to describe RIS-based algorithms, we need the following definitions. 
	Given a graph $G$ and edge influence probabilities $p_{uv}$ for every edge $e=(u,v)\in E$, one can generate an instance graph $g$ from $G$, by removing every edge $e=(u,v)$ by probability $1-p_{uv}$. Let $v$ be a node in graph $G$, then \textbf{reverse reachable (RR) set} for $v$ in $g$ is defined as the set of all nodes in $g$ that can reach $v$.
	An RR set is defined to be a \textbf{random RR set} if instance graph $g$ is created from $G$ as described before, and node $v$ is selected randomly from the set of nodes $V$.

In other words, every node $u$ that is present in an RR set generated for a particular node $v$, has an opportunity to activate $v$, if it has been activated already. Borgs et al. \cite{borgs2014maximizing} showed that the probability by which a node $u$ can appear in a random RR set is proportional to the expected influence spread of "that" node. Moreover, they proved that the probability by which a particular seed set $S$, overlaps a random RR set can be used as an unbiased estimator of expected influence spread of the seed set, as given in the following equation:

\begin{equation}
\mathbf{E} [ I(S) ] = n \times Pr[S \cap RR \neq \emptyset].
\label{eq:ris}
\end{equation}

\begin{algorithm}[tp]
	\begin{algorithmic}[1]
		\Input $G$, $k$
		\Output $S$
		\State $S = \emptyset, R = \emptyset$
		
		\State // Step 1: estimation and sampling
		\State Estimate $\theta$, i.e., the required number of RR sets 
		\State Generate $\theta$ random RR sets and insert them into R
		
		\State // Step 2: maximum coverage
		\For{$i = 1$ to $k$}
		\State $v =  \displaystyle \argmax_{u \in V\setminus S} \{ \sum_{RR\in R}\mathds{1}(u \in RR)  \}$
		\State $S = S \cup \{v\}$
		\State Remove all RR sets covered by $v$ from $R$
		\EndFor
	\end{algorithmic}
	\caption{RIS-based algorithm for the IM problem. } 
	\label{alg:RIS}
\end{algorithm}

According to above results, Borgs et al. \cite{borgs2014maximizing} proposed an algorithm to solve the IM problem. As illustrated in \mRefAlg{alg:RIS}, it consists of two main steps. First, given $\epsilon$, the required number of random RR sets $\theta$ to guarantee the approximation ratio $(1-\frac{1}{e}-\epsilon)$ is estimated, and then, a total number of $\theta$ random RR sets are generated. Second, the problem of maximum coverage on all generated random RR sets is considered. The goal of the problem is to find a set of $k$ nodes which covers the maximum number of RR sets. The problem is known to be NP-Hard, and it has been proven that the hill-climbing greedy algorithm yields a $(1-\frac{1}{e})$ approximate solution. After solving the maximum coverage problem, the resulting set of nodes (of size $k$) is returned as the solution to the IM problem \cite{borgs2014maximizing}.

Many previous works on RIS-based influence maximization algorithms have focused on providing formal analysis on the required number of random RR sets to ensure $(1-\frac{1}{e}-\epsilon)$-approximation guarantee \cite{tang2015influence, tang2014influence, nguyen2016stop}. Specifically, most studies have tried to reduce the number of required random RR sets, while preserving theoretical guarantees. 
Borgs et al. \cite{borgs2014maximizing} terminated the random RR set generation procedure when the total number of traversed edges during random RR set generation exceeds a predefined threshold $\tau$. They showed that by setting $\tau$ equal to $O(\epsilon^{-2}k(m+n)\log(n))$, a $(1-\frac{1}{e}-\epsilon)$-approximation solution is guaranteed. The main issue with this method is the large constant factor in the equation that determines the threshold, which incurs heavy computational cost.  
Tang et al. \cite{tang2014influence} proposed TIM and TIM+ algorithms. They showed that $\theta$ should be greater than $\frac{\lambda}{OPT}$, where parameter $\lambda$ is a constant factor determined by $n$, $\epsilon$ and $k$, and parameter $OPT$ is the optimal solution's expected influence spread. 
Later, Tang et al. proposed IMM \cite{tang2015influence}, which reduces the value of $\lambda$, and as a result, the value of $\theta$ based on a martingale approach \cite{williams1991probability}. 
Nguyen et al. \cite{nguyen2016stop} proposed SSA/DSSA algorithms and reduced $\theta$, but later, Huang et al. \cite{huang2017revisiting} showed that there are some issues with SSA's analysis and it cannot preserve theoretical guarantees.

Both TIM and IMM algorithms need optimal solution's influence spread, i.e., $OPT$, in order to estimate the required number of RR sets, but in reality there is no prior knowledge on the value of $OPT$. These two algorithms resolve this issue by providing a (preferably close) lower bound $LB$ for the value of $OPT$, and therefore, a larger estimated value for  $\theta$, i.e., $\theta = \frac{\lambda}{LB} > \frac{\lambda}{OPT}$.

\begin{algorithm}[tp]
	\begin{algorithmic}[1]
		\Input $G$, $k$, $\epsilon$ 
		\Output $LB$

		\State $S = \emptyset, R = \emptyset$
		
		\For{$i = 1$ to $\log_2(n) -1$}
		\State $x = n/2^i$
		\State $\theta = f(n, \epsilon, k) / x$
		\State Sample random RR sets until $|R| < \theta$
		\State $S = $ SeedSelection($R$, $k$)
		
		\If {$n\times F_R(S) \geq (1 + \sqrt{2}\epsilon)\times x$}
		\State $LB = n\times F_R(S)/(1+\sqrt{2}\epsilon)$
		\State break
		\EndIf
		
		\EndFor
	\end{algorithmic}
	\caption{Estimating the lower bound LB.} 
	\label{alg:estimation}
\end{algorithm}

To obtain a reasonable lower bound for $OPT$, both TIM and IMM use bootstrapping estimation techniques to perform a hypothesis testing (see \mRefAlg{alg:estimation}). They first set the value of $\theta$ with an initial estimation (line $4$). We refer readers to \cite{tang2014influence, tang2015influence} for more details on the function $f$ in line $4$. Next, they sample random RR sets until the total number of RR sets reaches $\theta$ (line $5$). Then, they construct a seed set from the generated RR sets. The function SeedSelection in line $6$ in \mRefAlg{alg:estimation} is equivalent to some extent to lines $6-10$ in \mRefAlg{alg:RIS}. After that, they estimate the influence spread of the generated seed set and compare it with the approximation bound of the estimator (line $7$). If the estimated influence spread is much higher than the approximation bound, they use a discounted value of that estimate as $LB$ (line $8$). Otherwise, they double the number of RR sets (lines $3-4$) and repeat the described procedure.

\section{Proposed RIS-Based Parallel Algorithm}
\label{sec:alg}

In this section, we present our solution to accelerate computational bottlenecks of RIS-based IM algorithms. Our GPU-accelerated IM algorithm is named gIM. Similar to IMM, our gIM algorithm is composed of two main steps: 1) sampling RR sets, and 2) generating the seed set.

\subsection{Baseline Parallel Methods for RR Set Sampling}
\label{sec:alg:ineff}

Before discussing our proposed method, we first investigate two simpler approaches that can be used to parallelize random RR set generation in RIS-based algorithms. For each of these two approaches, we first explain the method and then mention its drawbacks. Our proposed method for RR set sampling will be discussed in Section \ref{sec:alg:rr}.

\subsubsection*{Thread-Level Parallelization:}

Since in RIS-based algorithms, every random RR set can be generated independently from other random RR sets, one simple approach to parallelize random RR set generation is to assign the task of creating every random RR set to one thread, and launch a large number of threads on GPU. 

The main issue with this approach is its severe workload imbalance. Since size of a random RR set may range from a single node to a large portion of the graph, and node degrees vary as well, the amount of work to generate a random RR set varies drastically. This causes severe workload imbalance among the threads within a block, which makes a large number of threads inactive most of the time, and hence, highly degrades the overall performance.

Memory capacity is another impediment to this approach. Since every thread needs to separately maintain large arrays (e.g., visited nodes), memory capacity can easily become the bottleneck and prevent this approach to scale-up to large graphs with millions of nodes.

\subsubsection*{Parallel Breadth First Search:}

Generating a random RR set is equivalent to running breadth-first search (BFS) algorithm starting from a randomly selected node on a reversed instance graph. One can avoid generating a whole reversed instance graph by incorporating instance graph generation into the BFS algorithm. In the standard BFS, once a node $u$ is picked up from the front of the frontier queue, all of its outgoing edges are traversed. If instead of traversing all the outgoing edges, one traverses every outgoing edge of $u$ with probability of $p_{uv}$, then the result is equivalent to running a BFS on a reversed instance graph.

Therefore, one approach towards accelerating random RR set generation is to employ parallel algorithms previously proposed for BFS on GPU \cite{busato2014bfs, liu2015enterprise, merrill2012scalable}. Such methods parallelize the procedure of generating one random RR set on GPU. 

Most previous solutions on parallelizing BFS on GPU are level-synchronous, that is each level can be processed in parallel, but consecutive levels must be processed sequentially. These methods often maintain a node frontier queue, which is produced by processing the previous level, and is used to produce the node frontier for the next level. In level-synchronous methods, parallelization is usually implemented among the nodes in the current frontier queue or their outgoing edges, therefore, if the size of current frontier is relatively small, then not much gain can be attained from parallelization. Even the overhead of synchronization between consecutive levels can exceed any gain achieved from parallelization. Unfortunately, the problem of small frontier is not rare. Under the LT model, each node has at most one active outgoing edge so the maximum number of nodes in the frontier does not exceed $1$. Even under the IC model, where nodes may have more than one active edge, low values of influence probabilities in real data cause instance graphs to be usually much sparser than $G$. Therefore, the average number of active edges per node is low and the frontier cannot grow very much.

Gaihre et al. \cite{gaihre2019xbfs} proposed XBFS and implemented asynchronous BFS which allows nodes from different levels be visited in the same iteration. However, XBFS performs only one BFS at a time. 
Liu et al.\cite{liu2016ibfs} exploited the similarity between frontiers of BFS traversals started from different nodes and proposed an algorithm to run multiple concurrent BFSs on the same GPU. This algorithm still requires relatively large frontiers to achieve performance gains and is also heavily reliant on the assumption of high resemblance between frontiers, which may not be the case for generating random RR sets.

Another important issue with using previous parallel BFS algorithms is producing redundant nodes in the frontier queue. Since all nodes in the current frontier queue are processed in parallel, a node which is adjacent to more than one node in the current frontier might be placed in the next level's frontier more than once. In standard BFS, this issue may degrade speed, but does not affect the correctness of the algorithm. However, edge traversal in random RR set generation is randomized, and when a node is placed more than once in the frontier, its outgoing edges are processed multiple times, and this issue violates the correctness of RR set generation. For example, if node $u$ is placed in the frontier queue twice, each one of its outgoing edges $e=(u,v)$ are also processed twice, which accordingly increases influence probability from $p_{uv}$ to $1-(1-p_{uv})^2$.

\subsection{Proposed Parallel Method for RR Set Sampling}
\label{sec:alg:rr}

The above two approaches are actually located at two extreme ends of the spectrum. One launches many random RR set generation tasks but executes each one of them sequentially, and the other generates a single random RR set at a time and strives to parallelize that single task using all the GPU resources. 
gIM employs a parallel method which is located somewhere between these two extremes. 

\subsubsection*{Overview:}

We run many random RR set generations in parallel, and also, parallelize the execution of every random RR set generation to some extent. 
In specific, we judiciously assign the task of generating every random RR set to one CUDA block. See \mRefAlg{alg:rrpar}. 
Every block is associated with a fixed-size frontier queue in the shared memory which is denoted by $Q_{shr}$, a $Visited$ array of length $n$ to keep account of visited nodes during BFS, and a custom-designed data structure named $RR_{tmp}$ which is used to temporarily hold the generated RR set. Both $Visited$ and $RR_{tmp}$ are located in GPU global memory. 

The number of RR sets to which a particular node belongs, is an estimator of that node's influence spread. Hence, we also need to maintain the number of occurrences of every node in all the generated RR sets. To do so, we employ array $Occur$ of length $n$, whose elements are  initialized to zero. 

\subsubsection*{Data Structure $RR_{tmp}$:}

The number of nodes in an RR set may vary from one to all the nodes. Therefore, $RR_{tmp}$ should be capable of storing an RR set as large as the set $V$. If $RR_{tmp}$ is implemented as an array of length $n$, since every block is associated with a $RR_{tmp}$, the total number of blocks that can be executed in parallel would be restricted by the amount of memory required to store all these large arrays. If $RR_{tmp}$ is implemented as a linked list, whose memory footprint can grow dynamically, the overhead of memory management would be too large. 
To tackle this issue, we devise a data structure that is similar to a linked list. However, the granularity for memory management is set  larger than a single node. In specific, each element in the linked list is composed of a pointer to the next element and a fixed-length array to store some of the nodes.

\subsubsection*{Graph Representation:}
\begin{figure}[tp]
	\centering
	\includegraphics[width = 0.75\columnwidth]{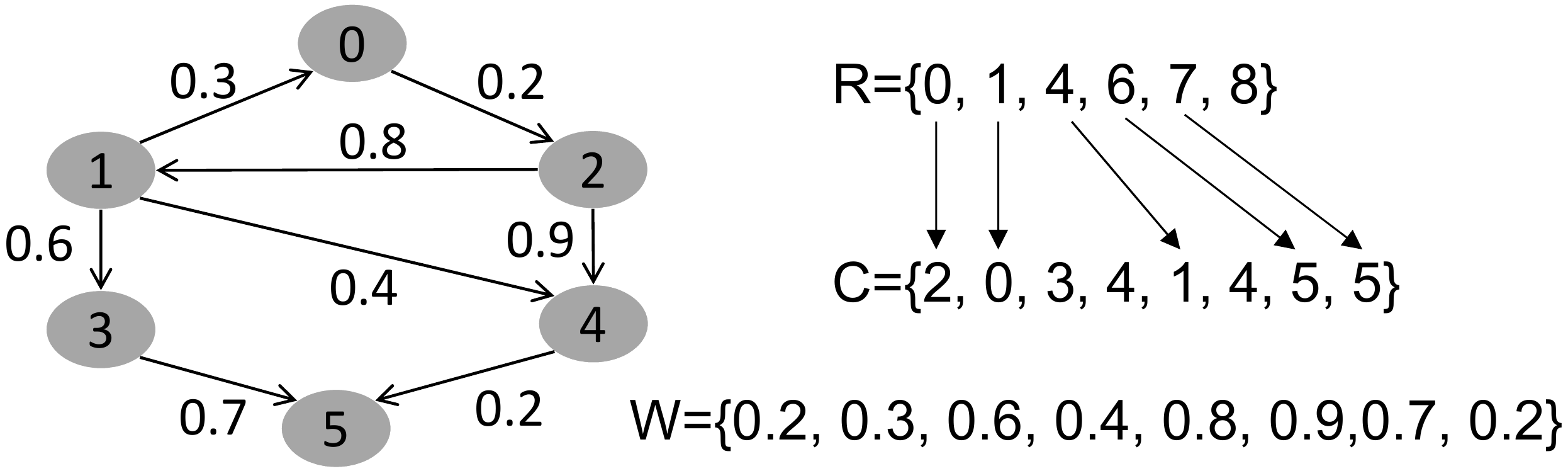}	
	\vskip -3mm
	\caption{Example: a graph and its corresponding CSR representation.} 
	\label{fig:csr_rep} 
\end{figure}

We represent graph $G$ in the compressed sparse row (CSR) format. As illustrated in \mRefFig{fig:csr_rep}, the CSR representation consists of two arrays, namely, the column indices array $C$, and the row offsets array $R$. For a graph $G$ with $n$ nodes and $m$ edges, array $C$ of size $m$ is created by concatenating adjacency lists of all the nodes in $G$. Array $R$ has $n+1$ elements, where element $R[i]$ denotes the location of the adjacency list of node $i$ in array $C$. One can find all outgoing neighbors of a particular node $i$, by reading all elements in $C$ indexed from $R[i]$ to $R[i+1]$. 
In addition, array $W$ of size $m$ contains influence probability $p_{uv}$ for every edge $e=(u,v)$. 
Note that in our CSR representation, all the nodes and edges are kept in the same ordering as $G$, i.e., we do not perform any sorting or pre-processing on the nodes or edges of the input graph $G$.

\begin{algorithm}[tp]
	\begin{algorithmic}[1]
		\Input $R, C, W, \theta$ 
		\Output $Occur, RR_{tmp}$
		\vskip 1mm
		\Block $\theta$
		\Thread $N_{th}$
		\vskip 1mm
		
		\If {$tx==0$}
		\State $RR_{tmp}.init()$
		\State $Q_{shr}.init()$
		\State $Q_{res}.init()$ ~~$\sslash$ \mRefSec{sec:alg:rr:overflow}
		\State $u = randSelect(V)$
		\State $Q_{shr}.enqueue(u)$
		\EndIf
		
		\While{$!Q_{shr}.empty()$}
		\State $u = Q_{shr}.front()$
		\If {$tx==0$}
		\State $RR_{tmp}.insert(u)$
		\State $Q_{shr}.dequeue()$
		\State $atomicAdd(Occur[u] ,1)$
		\EndIf
		
		\State $s = R[u]$, $e = R[u+1]$
		\For{$~~(i=tx;~~ i <e-s ;~~ i=i+N_{th})$} 
		\State	$offloadQueue(Q_{shr}, Q_{res})$ ~~$\sslash$ \mRefSec{sec:alg:rr:overflow}
		\State $v = C[s+i]$, $p_{uv}=W[s+i]$, $p = U(0, 1)$
		
		\If {$p < p_{uv}$ \& $Visited[v]==false$}
		\State $Visited[v] = true$
		\State $Q_{shr}.atomic\_enqueue(v)$
		\EndIf
		\EndFor
		\State $reloadQueue(Q_{shr}, Q_{res})$ ~~$\sslash$ \mRefSec{sec:alg:rr:overflow}
		\EndWhile
		
	\end{algorithmic}
	\caption{Parallel kernel for random RR set generation (see \mRefSec{sec:alg:rr}).} 
	\label{alg:rrpar}
\end{algorithm}

\subsubsection*{Algorithm Details:}
\mRefAlg{alg:rrpar} shows our parallel algorithm for random RR set generation under IC diffusion model. The input graph is represented by $R$, $C$ and $W$ arrays. Parameter $\theta$ is the number of required random RR sets to be generated. 

The algorithm works as the following. Every thread within a block is distinguished by an index denoted as $tx$, which ranges from $0$ to $N_{th}-1$. First, thread zero ($tx=0$) performs the required initializations, randomly selects a node from $V$ as the source node for BFS, and inserts this node at the front of $Q_{shr}$ (lines $2-6$). 
Next, all $N_{th}$ threads start a parallel randomized BFS from the randomly selected node. In this parallel BFS, nodes are processed sequentially and parallelization is realized in evaluation of adjacent edges to the current node. The BFS continues as long as the shared queue $Q_{shr}$ is not empty (line $8$). 

If $Q_{shr}$ is not empty, all threads extract the node at the front of $Q_{shr}$ (line $9$). 
Next, thread zero ($tx=0$) inserts the extracted node to $RR_{tmp}$, removes it from $Q_{shr}$, and increments its corresponding element in array $Occur$ (lines $10-14$). Then, all threads within the block find the range of the extracted node's out-neighbors in array $C$ (line $15$). Next, they collaboratively read the out-neighbors of the current node from $C$ and their corresponding influence propagation probabilities from $W$ (line $18$). 
While processing an edge, every thread produces a random number from the uniform distribution $U(0,1)$, and compares the generated random number with influence probability of that edge, and based on this comparison, it decides whether to traverse that edge or not. When a thread decides to traverse a particular edge, it checks the visited flag of the destination node of that edge, and if the node has not been visited already, it is added to the frontier queue (lines $19-22$). Since threads within a block are executed in parallel, more than one thread may decide to add a node to $Q_{shr}$. Therefore adding a node to $Q_{shr}$ should be protected by atomic operations (line $21$). 
Details of $offloadQueue()$ in line $17$ and $reloadQueue()$ in line $24$ are discussed in \mRefSec{sec:alg:rr:overflow}.

\subsection{Avoiding Overflow of the Frontier Queue}
\label{sec:alg:rr:overflow}

Shared memory is one of the most limited resources in GPU. 
In the proposed parallel algorithm, every block uses shared memory to store its own fixed-size frontier queue. If we allocate small queues, the actual frontier may grow larger and cause the queue to overflow, in which case, the produced RR set is no longer valid. 
In order to guarantee that no queue overflow occurs, we could increase the queue size to the total number of nodes. However, by doing this, we are no longer able to maintain the queue in GPU shared memory. Moving the queue to GPU global memory incurs long memory access latencies. To resolve this issue, gIM employs the following solution. We define a threshold $\tau_{q}$ as 
\begin{equation}
\tau_{q} = |Q_{shr}| - N_{th},
\label{eq:q_tau}
\end{equation}
where $|Q_{shr}|$ is queue capacity (the maximum number of nodes that every queue may hold), and $N_{th}$ is the total number of threads within a block. When the number of nodes currently stored in the queue exceeds $\tau_q$, there is a chance for the queue to overflow. This is because each one of $N_{th}$ threads within the block processes one edge, and therefore, has a chance to add one new node to the queue. In the worst case, when all $N_{th}$ threads add a node to the queue, overflow occurs.

\begin{algorithm}[tp]
	\begin{algorithmic}[1]
		\Function{offloadQueue}{$Q_{shr}, Q_{res}$}
		\If {$Q_{shr}.size() > \tau_q$}
		\If {$tx == 0$}
		\State $ptr = Q_{res}.alloc()$
		\EndIf
		\State $ptr[tx] = Q_{shr}.deque\_block(tx)$
		\EndIf
		\EndFunction
	\end{algorithmic}
	\caption{$offloadQueue$ function} 
	\label{alg:offloadq}
\end{algorithm}

\begin{algorithm}[h]
	\begin{algorithmic}[1]
		\Function{reloadQueue}{$Q_{shr}, Q_{res}$}
		\If{$Q_{shr}.empty()$ \& $!Q_{res}.empty()$}
		\If {$tx == 0$}
		\State $ptr = Q_{res}.pop()$
		\EndIf
		\State $Q_{shr}.enqueue\_block(ptr[tx], tx)$
		\EndIf
		\EndFunction
	\end{algorithmic}
	\caption{$reloadQueue$ function} 
	\label{alg:reloadq}
\end{algorithm}

To resolve this issue, we associate a reservoir queue, to every block, which is denoted by $Q_{res}$ and is actually implemented as a stack whose elements are arrays of length $N_{th}$. The reservoir queue is stored in GPU global memory and its size may grow dynamically, hence, it does not have shared queue's limitations. Our approach to avoid possible overflows is the following. 
As shown in line $17$ in \mRefAlg{alg:rrpar}, $offloadQueue(~)$ function is executed before all threads process their incident edges. 
As shown in \mRefAlg{alg:offloadq}, this function checks the condition $Q_{shr}.size() > \tau_q$. If the condition is satisfied, thread zero allocates the required memory to expand the reservoir queue. Then all threads collaboratively move $N_{th}$ elements from the shared queue to the newly allocated space in reservoir queue. This reduces the number of nodes in the shared queue and prevents possible overflow.

The elements of $Q_{shr}$ which have been moved to $Q_{res}$, need to be returned back to $Q_{shr}$ when it has enough empty space. This procedure is implemented in $reloadQueue()$ function, which is shown in \mRefAlg{alg:reloadq}. In this function, threads check if $Q_{shr}$ is empty and also if there are some nodes in the $Q_{res}$ that can be brought back to $Q_{shr}$ (line $2$). If this condition is satisfied, threads collaboratively move $N_{th}$ nodes from $Q_{res}$ to $Q_{shr}$ (line $6$). Once a batch of nodes are transferred from $Q_{shr}$ to $Q_{res}$, the order in which nodes are processed differs from the conventional BFS. However, this does not break the correctness of the parallel algorithm, as the order in which nodes are visited is not important.

\begin{figure}[tp]
	\centering
	\includegraphics[width = \columnwidth]{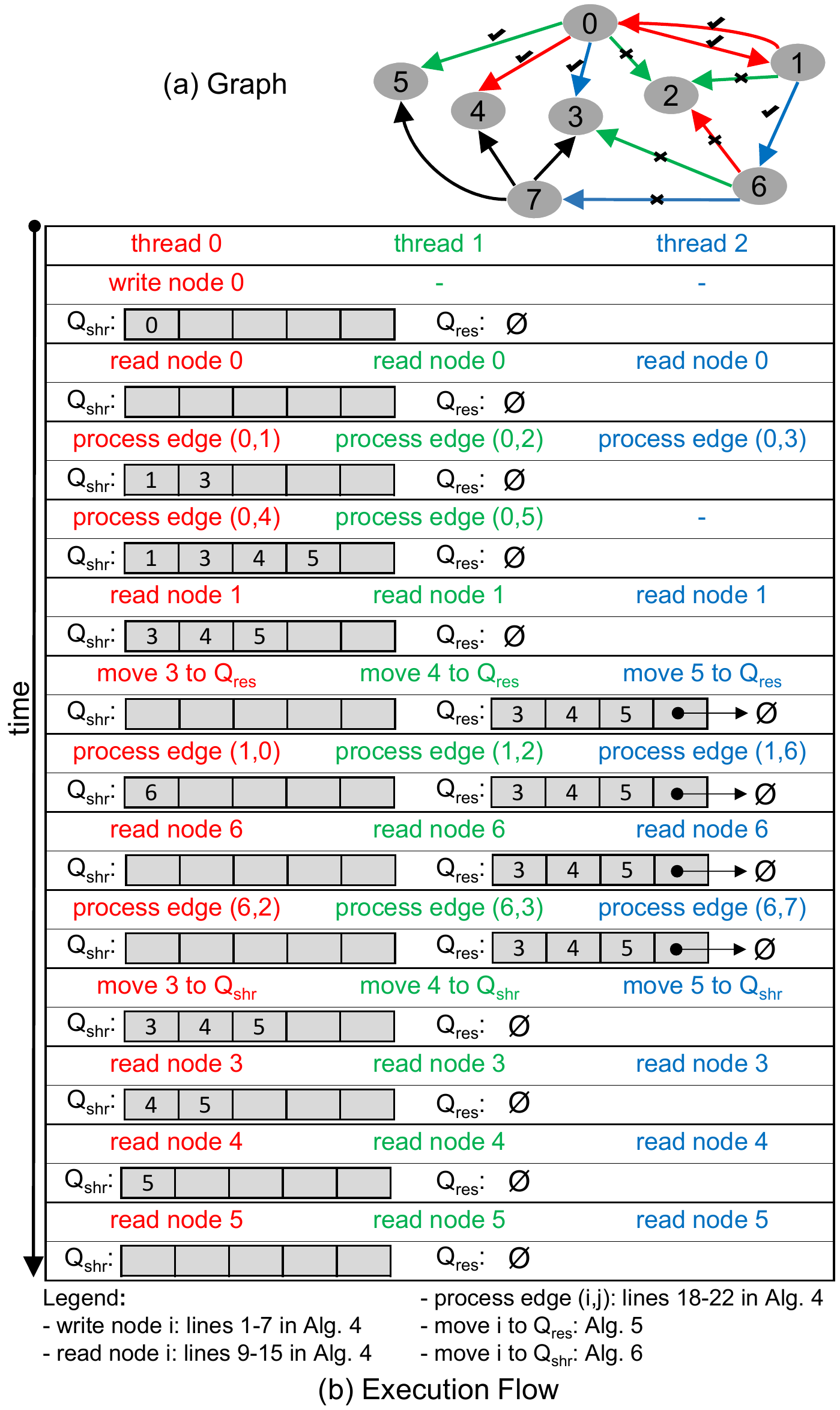}
	\vskip -4mm
	\caption{(a) An example graph. (b) Execution flow of a random RR set generation starting from node 0 within a small block with three threads. Details are discussed in \mRefSec{sec:alg:example}.}
	\label{fig:alg:example}	
\end{figure}

\subsection{An Illustrative Example}
\label{sec:alg:example}

\mRefFig{fig:alg:example} illustrates the execution flow of a random RR set generation a simple graph by using one small CUDA block with $N_{th}=3$ threads. Since outgoing edges are processed via parallel threads, edges are marked with different colors according to their corresponding threads. Black edges are not processed throughout this example, because their source node is not visited and thus is not placed in the queue. Check marks and cross marks on the edges represent whether the corresponding edge is traversed or not.

The execution starts by thread zero which places node $0$ into $Q_{shr}$. Next, node $0$ is extracted by all threads and its outgoing edges are processed. Since node $0$ has more than $N_{th}=3$ outgoing edges, processing these edges are performed in two consecutive steps and newly visited nodes are put in $Q_{shr}$. Then, node $1$ is extracted from  $Q_{shr}$, but threads find that $N_q=3 > \tau_q=2$, so $N_{th}$ nodes are moved from  $Q_{shr}$ to  $Q_{res}$. Next, outgoing edges of node $1$ are processed, and recently visited node $6$ is added to  $Q_{shr}$. In the next step, node $6$ is extracted from  $Q_{shr}$ and its outgoing edges are processed. Afterwards, threads find that  $Q_{shr}$ is empty and thus bring back $N_{th}$ nodes from  $Q_{res}$ to  $Q_{shr}$. The rest of the procedure is straightforward. Nodes are removed from the queue one by one until both  $Q_{shr}$ and  $Q_{res}$ become empty.

\subsection{Optimizing Memory Usage}
\label{sec:alg:rr:n_blocks}

In most cases, size of the generated RR set is very small, and therefore, $RR_{tmp}$ data structure has only few elements and the allocated array inside the last element of this data structure is underutilized. 
This under-utilization of GPU global memory should be avoided since the number of RR sets to be generated is large. 

To this end, \mRefAlg{alg:rrpar} is modified as shown in \mRefAlg{alg:rrpar_modified}. 
Right after producing an RR set, it is copied from $RR_{tmp}$ to a continuous chunk of memory called $RR$. This data structure is basically formed by concatenation of all the generated RR sets. 

In addition, the total number of concurrent blocks is reduced from $\theta$ (the number of RR sets to be generated) to a constant number $N_b$. Parameter $N_b$ is discussed later in \mRefSec{sec:alg:opt}. This is done because every block requires to store intermediate data (i.e., $Q_{res}$ and $Visited$) in GPU global memory.

In \mRefAlg{alg:rrpar_modified}, $N_{RR}$ and $tail_{RR}$ are scalar variables which lie in GPU global memory and keep track of the number of generated random RR sets so far, and the sum of their sizes, respectively. 
Once a random RR set is generated (line $2$), the location where it should be copied to in array $RR$ is determined and stored in $Offsets_{RR}$ (lines $4-5$). This is required so that the boundaries between consecutive random RR sets in array $RR$ can be distinguished later. In addition, $N_{RR}$ is incremented by $1$, and $tail_{RR}$ is incremented by the size of the generated RR set (lines $6-7$). 
Finally, all threads within the block collaboratively copy the generated RR set from $RR_{tmp}$ to the specified location in array $RR$ (lines $9-11$). 
A block terminates its execution when the number of generated RR sets has reached the number of required RR sets (line $12$).

Note that since $N_{RR}$ and $tail_{RR}$ are accessible by all concurrent blocks, lines $4-7$ should be protected from concurrent execution using mutex and atomicCAS in CUDA API. For brevity, this is not shown in \mRefAlg{alg:rrpar_modified}.

Without the optimization presented in this section, the required data would not fit into GPU global memory. For example, storing one $Visited$ array for a graph with $4$~million nodes requires about $488$~kB of GPU global memory, hence, if we were to launch $\theta$ blocks instead of $N_b$ blocks, launching $\theta=1$~million blocks would require $465$~GB of GPU global memory only to store all the $Visited$ arrays. This is about $10$ to $60$ times larger than the amount of memory available on even high-end GPUs today.

\begin{algorithm}[tp]
	\begin{algorithmic}[1]
		\Input $R, C, W, \theta$
		\Output $Occur, RR$ 
		\vskip 1mm
		\Block $N_b$ 
		\Thread $N_{th}$
		\vskip 1mm
		
		\Repeat		
		\State Generate one random RR set as in \mRefAlg{alg:rrpar}
		
		\If {$tx==0$}
			
			\State $offset = tail_{RR}$
			\State $Offsets_{RR}[N_{RR}] = offset$
			\State $N_{RR} = N_{RR} + 1$
			\State $tail_{RR} = tail_{RR} + RR_{tmp}.size()$

		\EndIf
		
		\For{$~~(i=tx;~~ i <RR_{tmp}.size() ;~~ i=i+N_{th})$} 
		\State $RR[offset + i] = RR_{tmp}[i]$
		\EndFor
		\Until{$N_{RR} \leq \theta$ }
	\end{algorithmic}
	\caption{Memory-optimized parallel kernel for random RR set generation (see \mRefSec{sec:alg:rr:n_blocks}).} 
	\label{alg:rrpar_modified}
\end{algorithm}

\subsection{Tuning to the Underlying Hardware}
\label{sec:alg:opt}

\subsubsection*{Tuning $N_{th}$ and Synchronization Mechanism:}

To ensure correct functionality, we need to synchronize all the threads within a block after lines $7$, $9$ and $14$ in \mRefAlg{alg:rrpar}, line $5$ in \mRefAlg{alg:offloadq} and line $5$ in \mRefAlg{alg:reloadq}. This is normally done by using $syncthreads()$ API call in CUDA. For example, a $syncthreads()$ call needs to be inserted after line $7$  because thread zero inserts a node in $Q_{shr}$, which affects the execution of other threads. 

The $syncthreads()$ API call is known to degrade performance. Hence, we propose the following optimization. 
In CUDA, every $32$ threads inside a block form a warp. All threads within a warp execute in lock steps. Hence, we set $N_{th}=32$, and as a result, no $syncthreads()$ calls are necessary anymore. This is because every block consists of only one warp, i.e., $N_{th}=32$ threads. Since all threads within a warp execute the same instruction, they are already in synchronization. 
This is the case in Nvidia Kepler, Maxwell, and Pascal GPU architectures. 

In recent Volta, Turing, and Ampere GPU architectures, however, the SIMT execution model has experienced some modifications, and as a result, the assumption of implicit synchronization among threads in a warp is no longer valid. Consequently, to ensure correct functionality, we need to insert $syncwarp()$ API calls after lines $7$, $9$ and $14$ in \mRefAlg{alg:rrpar}, and after line $5$ in both \mRefAlg{alg:offloadq} and \mRefAlg{alg:reloadq}. The $syncwarp()$ API call only synchronizes the threads within a warp, and has lower overhead compared to $syncthreads()$ API call. 

Therefore, we avoid $syncthreads()$ API call  by selecting $N_{th}=32$. 
Another reason for selecting $N_{th}=32$ is that average node degree is very low in many real-world graphs, and $32$ threads are enough to process all incident edges in parallel in most cases. Using larger values for $N_{th}$ causes a large number of threads to remain idle during the process of incident edges of low-degree nodes. See \mRefSec{sec:exp:bsize} for a discussion on the effect of larger values of $N_{th}$ on performance.

\subsubsection*{Tuning $N_b$ and $|Q_{shr}|$:}

In order to fully utilize parallel processing capacity of GPU, we determine the total number of blocks to be launched as the following equation:
\begin{equation}
N_b ~=~ \#SM \times MaxResBlockPerSM,
\label{eq:n_blocks}
\end{equation}
where $\#SM$ is the total number of streaming multiprocessors (SM) in the target GPU device, and $MaxResBlockPerSM$ is the maximum number of resident blocks per SM. 
The above formula means that we launch just enough blocks in \mRefAlg{alg:rrpar_modified} to keep all SMs busy, while avoiding the overhead of launching too many blocks, i.e., $\theta$ blocks as in \mRefAlg{alg:rrpar}.

To ensure that the number of concurrent blocks is not limited by the amount of available shared memory, $|Q_{shr}|$ is determined such that:
\begin{equation}
\lfloor \frac{|ShMem|}{|Q_{shr}|} \rfloor ~\geq~ MaxResBlockPerSM,
\label{eq:sh_q_cond}
\end{equation}
where $|ShMem|$ is the amount of shared memory per SM, and $|Q_{shr}|$ is the capacity of the fixed-size frontier queue. 
For a GPU with $|ShMem|=48$~KB and $MaxResBlockPerSM=32$, we must select $|Q_{shr}| < 384$, i.e., $1.5$~KB.

\vspace*{-5mm}
\subsection{Modifications under the LT Model}
\label{sec:alg:rr:lt}

Our proposed parallel algorithm for random RR set sampling can be modified to support LT diffusion model as well. 
Under this model, every node has at most one active incoming edge, which is selected randomly according to the edge weights. 
First, thread zero samples a random number from $U(0,1)$. Then all threads within the block run a parallel scan algorithm \cite{harris2007parallel} on the edge weights, and instead of comparing the generated random number with edge weights, they compare it with two consecutive numbers that resulted from the parallel scan. The first thread that finds the generated random number between two consecutive outputs of scan, marks its edge as active and broadcasts an early termination signal to all other threads within the block.

In addition, since every node visits at most one other node during random BFS, the size of the frontier queue never exceeds one, and therefore, the fixed-size shared frontier queue never overflows and the reservoir queue is not required anymore.

\begin{algorithm}[tp]
	\begin{algorithmic}[1]
		\Input $RR$, $Occur$, $u$
		\Output $Occur$, $Covered$
		\vskip 1mm
		\Block $N_b$
		\Thread $N_{th}$
		\vskip 1mm
		
		\For{$~~(i=bx;~~ i <N_{RR} ;~~ i=i+N_b)$} 
		\If {$Covered[i] == false$}
		\State $found = false$
		\State $offset = Offsets_{RR}[i]$
		\State $length = Offsets_{RR}[i+1] - offset$
		
		\For{$(j=tx;~ j<length ~\&~ !found;~ j=j+N_{th})$}
		\If { $RR[offset+j] == u$}
		\State $found = true$
		\EndIf						
		\EndFor				
		
		\If {$found == true$}			
		\State $Covered[i] = true$
		\For{$~~(j=tx;~~ j < length;~~ j=j+N_{th})$}
		\State $atomicAdd(Occur[RR[offset+j]], -1)$
		\EndFor
		\EndIf
		
		\EndIf
		\EndFor		
	\end{algorithmic}
	\caption{Maximum Coverage parallel kernel.}
	\label{alg:max_cov}
\end{algorithm}

\subsection{Proposed Parallel Method for Seed Set Generation}
\label{sec:alg:max_cover}

The final step is to solve the Maximum Coverage problem on the generated random RR sets, in order to generate the seed set. IMM \cite{tang2015influence} used the standard greedy algorithm to solve this problem, which is illustrated in lines $6-10$ of \mRefAlg{alg:RIS}.

Our proposed solution, however, employs a more efficient approach. As mentioned in \mRefSec{sec:alg:rr}, every node is associated with a counter which keeps track of the number of its occurrences across all the generated random RR sets. All such counters are stored in array $Occur$. 
By running a parallel reduction algorithm \cite{harris2007optimizing} with $\max$ operator on this array, the node with maximum coverage is found. 

Once a node is selected and added to the seed set, all RR sets which contain that node must be removed, and also, all counters whose corresponding nodes are present in those RR sets must be decremented. To do this, we associate every RR set with a $Covered$ flag which indicates whether or not the RR set is covered by the seed set so far. In order to mark the covered RR sets and update the counters, we employ the parallel kernel shown in \mRefAlg{alg:max_cov}. 
Here, the generated RR sets are distributed among $N_b$ blocks (line $1$). When a block processes an RR set, it first checks that the RR set has not already been covered by any node in the seed set (line $2$). Then, all threads within the block collaboratively search for all occurrences of node $u$, i.e., the newly added node in the seed set (lines $3-10$). If the RR set contains node $u$, it is flagged as covered, and all threads within the block collaboratively decrement the counters whose corresponding nodes are present in this RR set (lines $11-16$). 
Note that all the threads within a block need to be synchronized after lines $2$ and $10$, according to the discussion in \mRefSec{sec:alg:opt}.

\section{Experimental Evaluation}
\label{sec:exp}

\subsection{gIM Source Code}
\label{sec:exp:source}

The proposed gIM algorithm is implemented in C++ language using CUDA parallel programming framework. The source code of gIM is available online \cite{SRC}.

\subsection{Experiment Setup}
\label{sec:exp:setting}

Benchmark datasets are graphs extracted from real social networks. All of these datasets are publicly available for download from \cite{snapnets}, and have been widely employed in the literature on the IM problem to assess the influence spread and running time of different algorithms \cite{tang2015influence, minutoli2019fast, minutoli2020curipples, cohen2014sketch}. Their main characteristics are shown in Table~\ref{tab:datasetSpec}. 

We compare our proposed gIM algorithm with IMM \cite{tang2015influence},  Ripples \cite{minutoli2019fast}, cuRipples \cite{minutoli2020curipples} and SKIM \cite{cohen2014sketch}. IMM is the state-of-the-art algorithm which provides theoretical guarantee on the approximation ratio of influence spread. Ripples is a framework which provides multi-core implementation of IMM, and cuRipples is the extended version of Ripples that adds support for multi-GPU systems.  SKIM is among the best algorithms which does not provide the theoretical guarantee but runs faster than IMM.

To run the experiments, we employed a Linux machine with Intel Xeon Scalable 4110 CPU operating at $2.50$~GHz, and Tesla V100 GPU operating at $1245$~MHz and equipped with 16GB of global memory. 
We employ CUDA version $10.2$. 
Since IMM \cite{tang2015influence} and SKIM \cite{cohen2014sketch} are sequential methods, they are executed on a single CPU core, while Ripples and cuRipples use all the available CPU cores. 
In gIM, the parallel kernels are executed on GPU, and the rest of the program is executed on a single CPU core.

Although some efforts \cite{saito2008prediction, goyal2010learning, kutzkov2013strip} have been made to learn influence probabilities based on user actions extracted from social networks, most previous works used heuristics to assign influence probabilities to the edges. Weighted Cascade (WC) \cite{kempe2003maximizing} is the most commonly used schemes in previous works \cite{chen2009efficient, tang2014influence, tang2015influence, cohen2014sketch}. In this scheme, the influence probability of edge $e=(u,v)$ is set to $1/d^{in}_v$, where $d^{in}_v$ is the in-degree of node $v$. By using this scheme, the probability of a node being influenced is heuristically less dependent on the number of incoming edges. Also, WC scheme, in which the sum of incoming edges' probabilities for each node is exactly equal to 1, can be used under the LT model. 
In our experiments, we use WC scheme to assign influence probabilities to edges, under both IC and LT models.

\setcounter{table}{0}
\begin{table}[tp]
	\centering
	\caption{Benchmark datasets.}
	\vskip -3mm
	\label{tab:datasetSpec}
	\begin{tabular}{|c|c|c|c|c|}
		\hline
		Dataset & Type&\# of nodes ($n$)	& \# of edges ($m$)  \\
		\hline
		soc-Epinions1& Directed & 75,879& 508,837\\
		\hline
		soc-Slashdot0922& Directed & 77,360 & 905,468\\
		\hline
		higgs-twitter& Directed & 456,631& 14,855,875	\\
		\hline
		soc-Pokec& Directed & 1,632,803& 30,622,564\\
		\hline		
		soc-LiveJournal1& Directed & 4,847,571 & 68,993,773\\
		\hline
		com-Orkut& Undirected & 3,072,441 & 117,185,083\\
		\hline		
	\end{tabular}
	\vskip -3mm
\end{table}

\setcounter{table}{1}
\begin{table*}[tp]
	\centering
	\vskip -2mm
	\caption{
		Runtime of different methods under IC diffusion model.
	}
	\vskip -3mm
	\label{tab:overall}
	\begin{tabular}{|c|c|c|c|c|c|c|c|}
		\hline
		\multicolumn{2}{|c|}{}	  & soc-Epinions1	& soc-Slashdot0922 & higgs-twitter & soc-Pokec & soc-LiveJournal1 & com-Orkut	\\	\hline
		
		& IMM \cite{tang2015influence} & 4.42 & 12.69 & 92.94	& 122.91 & 251.86 & 813.09 \\	\cline{2-8} 
		
		& Ripples \cite{minutoli2019fast}	& 11.09 & 18.35	& 56.86	& 65.23 & 133.71 & 586.52 \\	\cline{2-8}
		
		Runtime & cuRipples \cite{minutoli2020curipples}	& 1.25 & 1.68 & 13.08 & 21.42 & 87.53 & 96.98  \\	\cline{2-8}
		
		(sec.)  	& SKIM \cite{cohen2014sketch} 			& 0.29 & 0.58	& 2.49	& 4.43	& 7.71 & 36.53	\\	\cline{2-8}		
		
		& gIM (proposed)		& 0.131	& 0.143	& 0.802	& 1.084	& 2.276	& 3.689	\\	\hline 
		\hline
		
		& Ripples \cite{minutoli2019fast}& 0.40 & 0.69 & 1.63 & 1.88 & 1.88 & 1.39 \\	\cline{2-8}
		
		Speedup ratio& cuRipples \cite{minutoli2020curipples}& 3.54 & 7.55 & 7.11 & 5.74 & 2.88 & 8.38 \\	\cline{2-8}
		
		over IMM & SKIM \cite{cohen2014sketch}     & 15.24	& 21.87 & 37.33 & 27.74 & 32.67	& 7.82  \\	\cline{2-8}		
		& gIM (proposed)	 & 33.74 & 88.74 & 115.89 & 113.39 & 110.66 & 220.41	\\ 	\hline
	\end{tabular}
\end{table*}

\subsection{Performance Comparison}
\label{sec:exp:others}

Table~\ref{tab:overall} illustrates the runtime of different methods with $k=50$ and $\epsilon=0.05$ under the IC model in different datasets. 
In every algorithm, and for every dataset, the algorithm is executed ten times, runtime values are measured, and the average runtime is reported. The observed standard deviation is about $2\%$ of the average value. 
Note that the entire runtime is measured, which includes the time required to perform memory allocations on GPU and  transfer data between CPU and GPU.

We used the default parameters in the evaluation of SKIM. 
It is noteworthy to mention that the solution quality, i.e., the influence spread of the resulting seed set, is the same for IMM and gIM, however, the solution quality of SKIM is slightly lower (about $1\%$ to $3\%$ lower). 

The runtime of IMM varies from $4.42$ seconds to $813.09$ seconds, and it increases as graph size grows. 
SKIM runs faster than IMM. The runtime varies from $0.29$ to $36.53$ seconds. 
gIM, however, attains much smaller runtime compared to both IMM and SKIM. Our solution's runtime ranges from $131$ milliseconds to $3.689$ seconds. Note that the measured runtime values include all overheads such as CPU to GPU data transfer latency. The speedup ratio of gIM over IMM ranges from $33.74\times$ to $220.41\times$.

The runtime of Ripples' multi-core implementation of IMM, with $16$ parallel threads, ranges from $11.09$ to $586.52$ seconds. 
cuRipples has lower runtimes compared to Ripples. However, despite using the computational power of both CPU and GPU, cuRipples is slower than gIM, which only employs the computational power of GPU. 
cuRipples moves the generated RR sets back and forth between the device and host memories, and therefore, is not limited to the size of GPU memory. 
cuRipples used two different methods for GPU implementation of random BFS: under the IC model, it employed a parallel BFS which generates one RR set at a time, and under the LT model, it adopted thread-level parallelism on GPU. Both of these methods are very similar to what discussed in \mRefSec{sec:alg:ineff}.

It is noteworthy to mention that Minutoli et al. \cite{minutoli2019fast,minutoli2020curipples} used a uniform distribution $U(0,1)$ to assign weights to the edges, while similar to other previous works such as \cite{chen2009efficient, tang2014influence, tang2015influence, cohen2014sketch} we use WC scheme. 
By using uniform distribution, half of the edges become active during the generation of a random RR set, because the mean of $U(0,1)$ is $0.5$. 
While in WC scheme, since the in-degree of most nodes is larger than two, most edges are assigned with weights less than $0.5$, and therefore, we expect to encounter less active edges during the generation of a random RR set in comparison with the uniform distribution model. 
Therefore, the difference between our experimental results for cuRipples and the ones in \cite{minutoli2020curipples} is due to the usage of different methods for assigning weights.

\subsection{Runtime Breakdown}
\label{sec:exp:breakdown}

\begin{figure}[tp]
	\centering
	\includegraphics[height = 35mm]{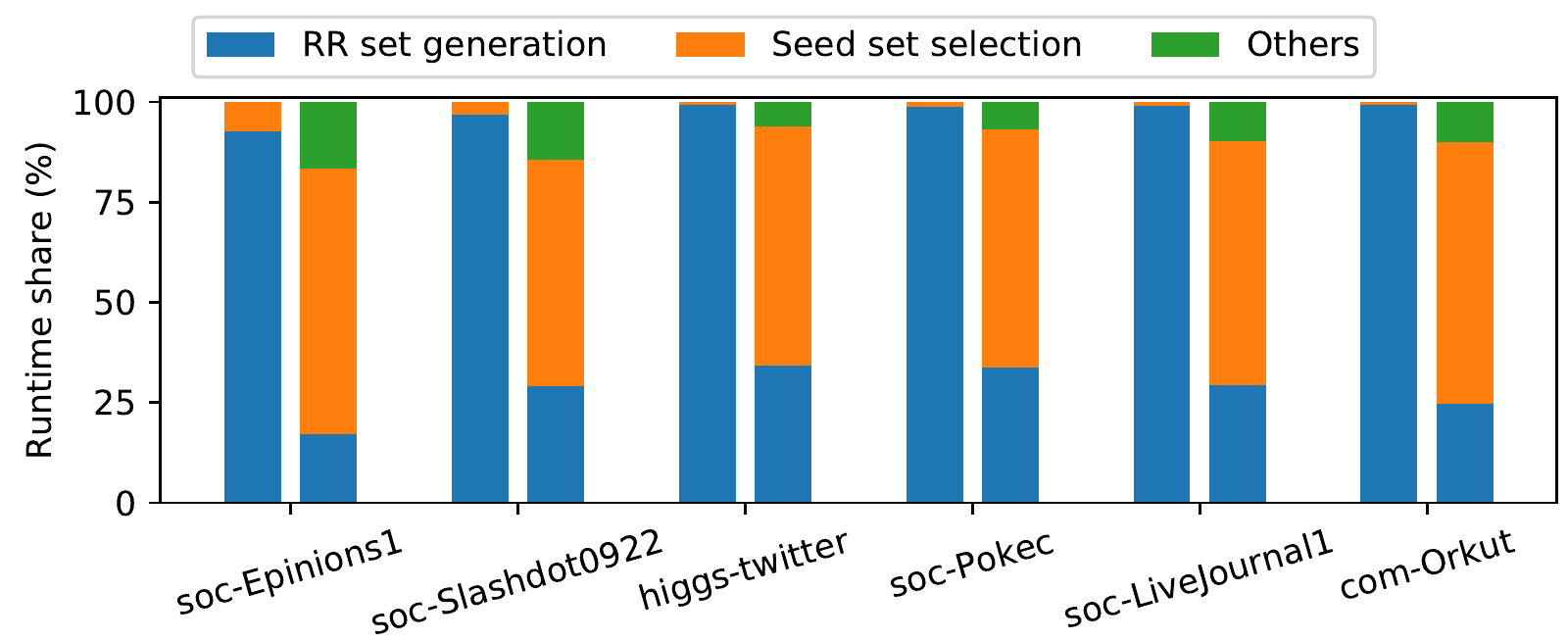}
	\vskip -4mm	
	\caption{Runtime breakdown of gIM (the proposed solution) compared to IMM. In each dataset, the left bar represents IMM and the right bar gIM.}
	\label{fig:perfbreak} 
\end{figure}

RIS-based algorithms can be decomposed into two main parts: random RR set sampling and seed selection, each of which has different computational characteristics. \mRefFig{fig:perfbreak} illustrates the share of each part in the total runtime, for both IMM and gIM. Note that the time required for memory allocation and moving data between CPU and GPU are also taken into account in gIM results. 

In IMM, a large portion of the runtime is spent on generation of random RR sets, while in gIM, seed set selection has a larger share of the runtime. This is because the speed-up ratio for our GPU acceleration is different for the two parts, in specific, the speedup ratio is higher for random RR set sampling in comparison with seed set selection.

\begin{figure*}[tp]
	\centering
	\vskip -1mm	
	\includegraphics[width = \textwidth]{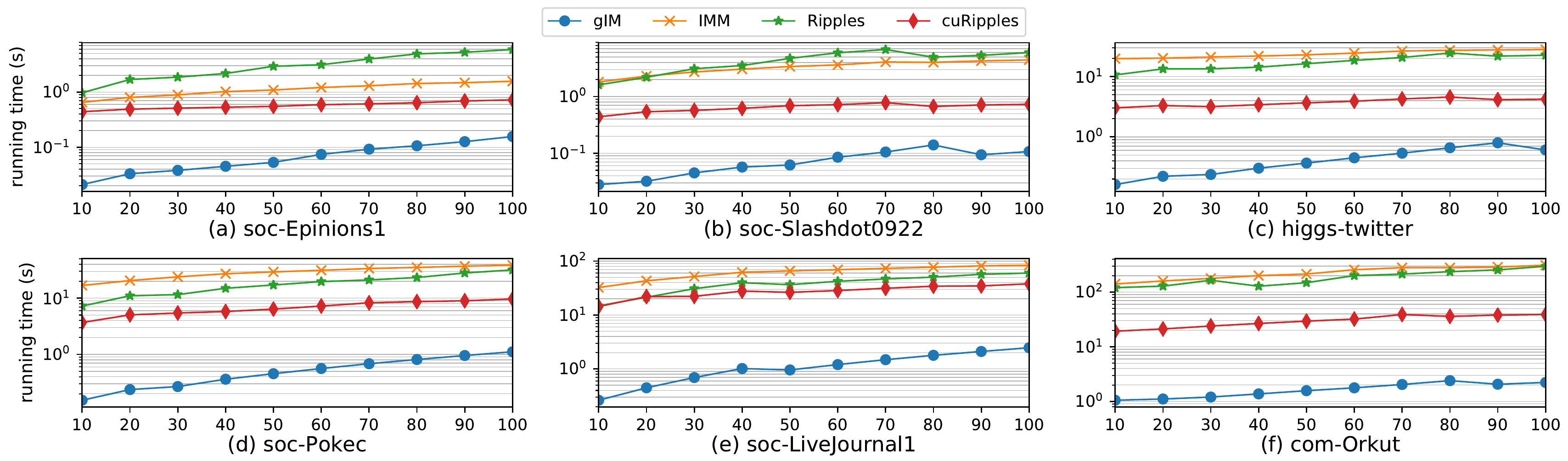}
	\vskip -4mm
	\caption{Runtime of different algorithms (y axis) versus $k$ (x axis) under IC model and $\epsilon=0.1$.}
	\label{fig:exp:ic_runtimes}	
\end{figure*}

\begin{figure*}[tp]
	\centering
	\vskip -3mm
	\includegraphics[width = \textwidth]{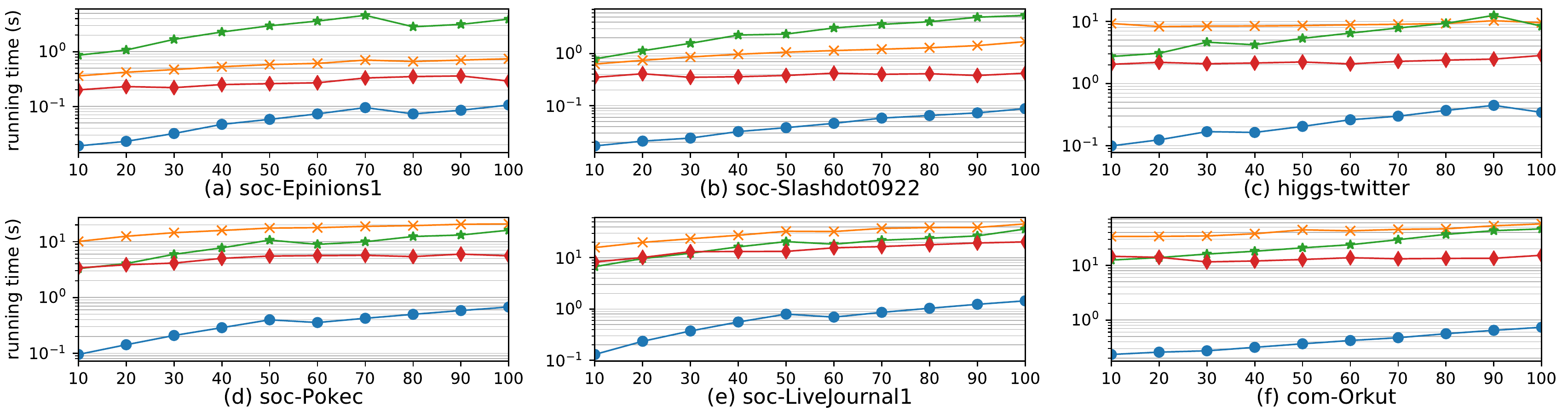}
	\vskip -4mm	
	\caption{Runtime of different algorithms (y axis) versus $k$ (x axis) under LT model and $\epsilon=0.1$.}
	\label{fig:exp:lt_runtimes}	
\end{figure*}

\begin{figure*}[tp]
	\centering
	\vskip -3mm
	\includegraphics[width = \textwidth]{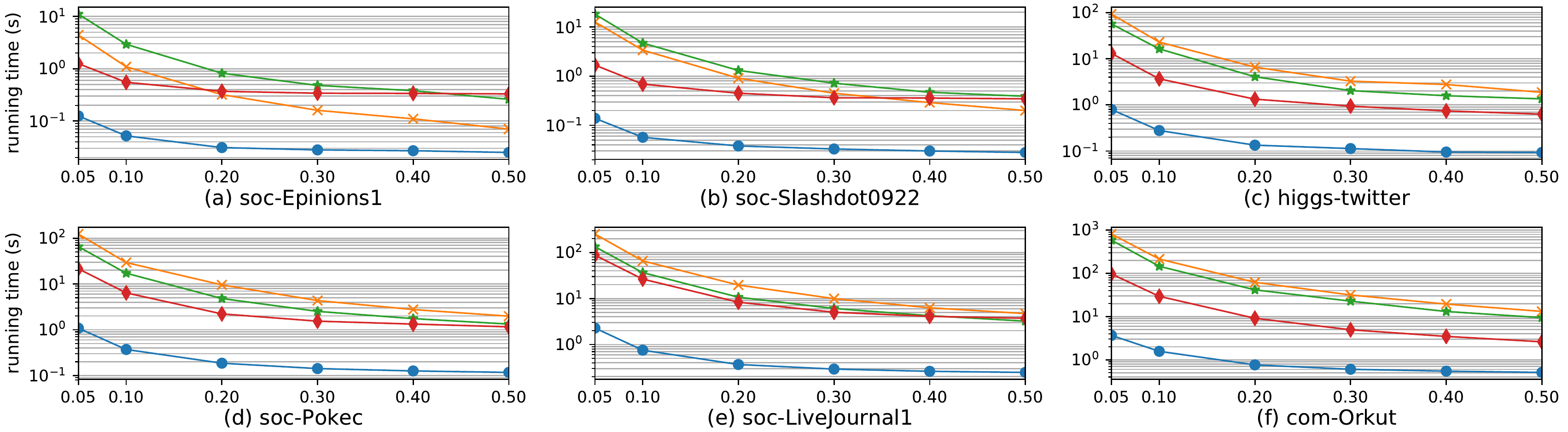}
	\vskip -4mm
	\caption{Runtime of different algorithms (y axis) versus $\epsilon$ (x axis) under IC model and $k=50$.}
	\label{fig:exp:eps_effect}	
\end{figure*}

\subsection{Impact of Parameters $k$ and $\epsilon$}
\label{sec:exp:params}

The number of required random RR sets to guarantee $(1-\frac{1}{e}-\epsilon)$-approximation ratio, which directly affects runtime, is dependent on the values of $k$ and $\epsilon$. \mRefFig{fig:exp:ic_runtimes} and \mRefFig{fig:exp:lt_runtimes} compare the runtime of gIM against IMM with $\epsilon=0.1$ and for different values of $k$. As can be seen, the speedup factor is relatively consistent for different values of $k$. It can be seen that the runtime of IMM is monotonically increasing, while the runtime of gIM sometimes drops when $k$ increases. 
This is because by increasing the value of $k$, the amount of influence of the seed set also increases and it might cause the condition in line $7$ of \mRefAlg{alg:estimation} to be satisfied one iteration earlier. In IMM, calling greedy seed set generation procedure for one fewer iteration does not significantly affect the runtime, because as it is evident in \mRefFig{fig:perfbreak}, random RR set generation is the dominant contributor to the runtime. While in gIM, random RR generation experiences much higher speedup in comparison with greedy seed selection. Therefore, the execution time of greedy seed selection procedure calls become comparable to that of random RR set generation, and one fewer iteration to this procedure can decrease the runtime in spite of the increased required number of RR sets.

We also studied the impact of $\epsilon$ on runtime. \mRefFig{fig:exp:eps_effect} illustrates the running times for different values of $\epsilon$  with $k=50$. The number of required RR samples (i.e., $\theta$) has an inverse quadratic relation with the value of $\epsilon$.  As can be seen, the runtime values follow the same trend, and therefore the amount of speedup is almost preserved for different values of $\epsilon$.

\subsection{Scalability}
\label{sec:exp:scal}

In this section, we evaluate the scalability of gIM. Specifically, we measure the runtime of gIM against IMM on synthetic graphs with varying densities and show that the speedup increases, as the density of the input graph grows.

In order to generate synthetic graphs, we use Barabasi-Albert model, which is an algorithm for generating undirected random scale-free graphs by using preferential attachment mechanism. This algorithm starts with a clique of $r_0$ nodes, and then, new nodes are added one by one. Every newly added node is randomly connected to $r$ nodes that are already in the graph ($r \leq r_0$). The constant parameter $r$ is the graph density. The probability that the newly added node is connected to node $i$ is determined by $p_i = \frac{d_i}{\sum_{j}d_j}$, where $d_i$ denotes the current degree of node $i$, and the sum is over the nodes that already exist in the graph. It has been proven that Barabasi-Albert model generates a graph with a power law degree distribution \cite{barabasi1999mean}.

\begin{figure}[tp]
	\centering
	\includegraphics[width = 0.9\columnwidth]{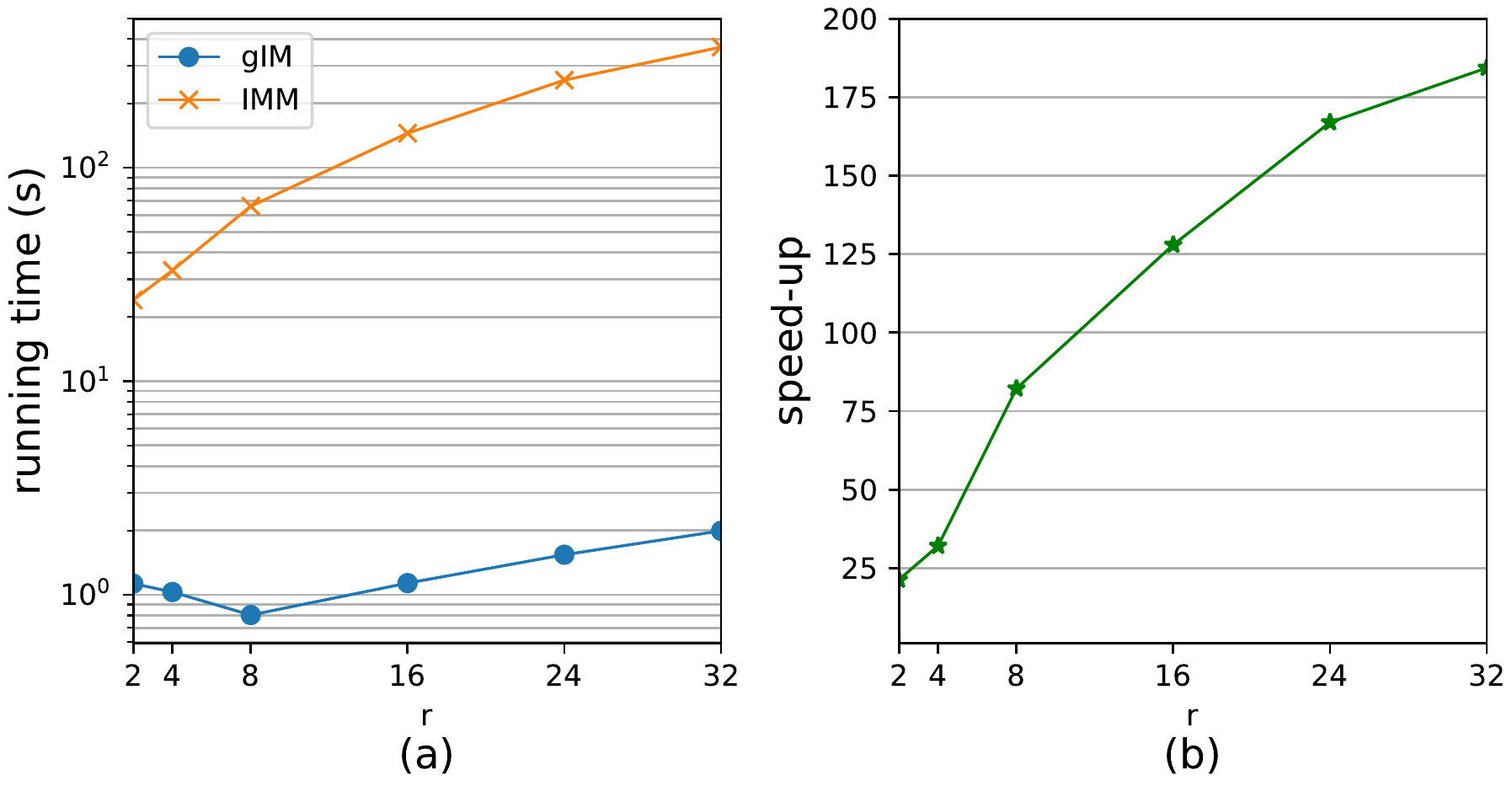}	
	\vskip -4mm
	\caption{(a) Runtime of gIM (the proposed solution) and IMM on random scale-free graphs with different densities. (b) The resulting speedup, i.e., the ratio of the two curves in (a).}
	\label{fig:scalability} 
\end{figure}

We generated various graphs with $n=10^6$ nodes and different densities ranging from $r=2$ to $r=32$, by using Barabasi-Albert algorithm. We measure the runtime of gIM and IMM over the generated graphs, with parameters $k=50$ and $\epsilon=0.05$. \mRefFig{fig:scalability} illustrates the resulting runtimes and speedup. As can be seen, the speedup increases, as $r$ grows. This observation can be attributed to the way by which edges are processed in gIM: $N_{th}$ threads within a warp process all outgoing edges of a single node. Therefore, if the number of outgoing edges of a node is less than $N_{th}$, some threads remain idle and perform no useful computation. When $r$ grows, the average degree also increases, and as a result, fewer threads remain idle and more potential parallelism can be exploited.

\vspace*{-3mm}
\subsection{Impact of Block and Grid Size}
\label{sec:exp:bsize}

\begin{figure}[tp]
	\centering
	\includegraphics[width = 1.0\columnwidth]{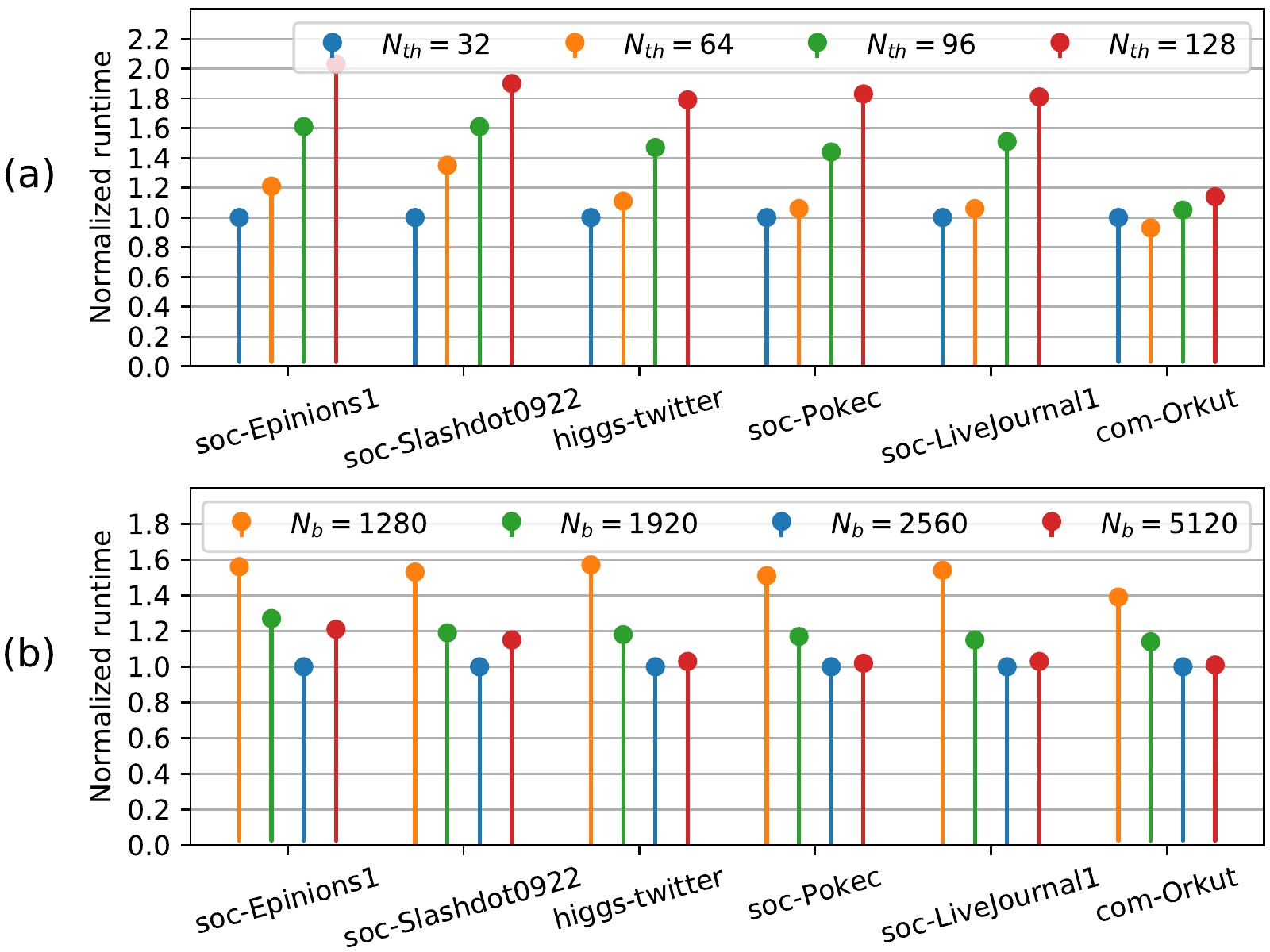}
	\vskip -3mm	
	\caption{(a) Runtime of gIM for $k=50$ and $\epsilon=0.05$, with different values of $N_{th}$, normalized to the runtime of $N_{th}=32$ (the selected block size). (b) Runtime of gIM for different values of $N_b$, normalized to the runtime of $N_b=2560$ (the selected grid size).}
	\label{fig:nth_impact} 
\end{figure}

As discussed in \mRefSec{sec:alg:opt}, we set the number of threads per block to $N_{th}=32$ to avoid having too many idle threads while processing low-degree nodes. In order to evaluate the impact of $N_{th}$ on the performance of gIM, we replaced $syncwarp()$ API calls with $syncthreads()$ API calls and set $N_{th}=64$, $96$, and $128$. \mRefFig{fig:nth_impact}a illustrates the runtime with different values of $N_{th}$, normalized to the runtime of $N_{th}=32$. As it can be seen, the performance degrades by increasing $N_{th}$. One reason is that the average degree in social graphs is relatively low, and having a larger block causes a higher percentage of the threads to remain idle while processing low-degree nodes. In addition, since the maximum number of resident threads per SM is constant, having larger blocks means fewer number of blocks may run  concurrently.

To efficiently utilize GPU parallel computing capabilities, we set the number of blocks, i.e grid size, according to \mRefEq{eq:n_blocks} to $N_b=2560$ for the employed V100 GPU device. In order to analyze the effect of grid size on performance, we varied $N_b$ and measured the runtime of gIM on different datasets. The resulting execution times are shown in \mRefFig{fig:nth_impact}b. As can be seen, the smallest execution times are achieved when $N_b$ is set to the value specified by \mRefEq{eq:n_blocks}. According to this equation, $N_b=2560$ blocks are enough to fully utilize the available hardware resources. Increasing $N_b$ results in more block management overheads, and hence, slightly increases the runtime. Reducing $N_b$ under-utilizes the GPU, and hence, increases the runtime.

\vspace*{-3mm}
\subsection{Multi-round Influence Maximization}
\label{sec:exp:MR}

Although some previously-proposed algorithms for solving the standard influence maximization problem have successfully been used to resolve some real-world problems \cite{yadav2016using, gibbs2017solving}, it is evident that the standard IM problem cannot grasp all aspects of real-world phenomena. Therefore, many researchers have attempted to propose new variations of the standard IM problem, in order to incorporate new details such as time restriction \cite{chen2012time}, location awareness \cite{li2014efficient} and topic awareness \cite{chen2016real} into the standard IM problem and make it more realistic for real-world situations. Interestingly, most of the variations to the IM problem can be easily solved by making some subtle modifications to RIS-based algorithms. Therefore, our parallel algorithm not only is able to solve the standard IM problem in a very short time but also is capable of solving different variations of it by applying simple modifications. In order to illustrate this, we introduce a specific variation of the IM problem called multi-round influence maximization \cite{sun2018multi} and show that our algorithm is able to solve this problem.

\setcounter{table}{2}
\begin{table}[tp]
	\centering
	\caption{Running time of CR-NAIMM algorithm in seconds.}
	\vskip -2mm
	\label{tab:MRIMM}
	\begin{tabular}{|c|c|c|c|c|}
		\hline
		Dataset & GPU& CPU & Speed-up  \\
		\hline
		soc-Epinions1& 0.087 & 2.4& 27.59\\
		\hline
		soc-Slashdot0922& 0.066 & 8.97 & 135.91\\
		\hline
		higgs-twitter& 0.74 & 56& 75.68	\\
		\hline
		soc-Pokec& 2.055 & 60& 29.13\\
		\hline		
		soc-LiveJournal1& 10.755 & 113 & 10.51\\
		\hline
		com-Orkut& 5.118 & 591 & 115.47\\
		\hline
	\end{tabular}
\end{table}

In the multi-round influence maximization problem (MRIM), influence propagates in multiple rounds, and the goal is to find a seed set for each round in order to maximize the total number of nodes that have been influenced at least once. Sun et al. \cite{sun2018multi} proposed three different algorithms to solve the MRIM problem, all of which can be implemented efficiently by using reverse influence sampling. We slightly modify our algorithm to parallelize CR-NAIMM algorithm in \cite{sun2018multi}, which is more computationally expensive than the others. Specifically, after selecting a random node, we initiate a random BFS originating from the selected node as many times as the number of rounds. Also, each element in a random RR set is a tuple of node-id and round number. The rest of our algorithm remains almost intact. We performed experiments using real-world datasets. We set the size of the seed set to $k=10$, the number of rounds to $T=5$, and $\epsilon=0.1$. The runtimes of both serial and parallel algorithms on different datasets are shown in Table \ref{tab:MRIMM}. We achieve a speed-up of up to $136 \times$ and an average of $65.72 \times$ on six datasets.

\section{Conclusion}
\label{sec:conc}

In this paper, we proposed gIM, an efficient parallel implementation of IMM algorithm on GPU. Extensive experiments on real social network graphs demonstrated that gIM significantly reduces the runtime of IMM. Moreover, we show that our algorithm is able to solve other variants of the IM problem, only by applying minor modifications.



\appendices


%

\ifCLASSOPTIONcaptionsoff
  \newpage
\fi



%

%

 \bibliographystyle{IEEEtran}
\bibliography{reference}
\vspace*{-6mm}
\begin{IEEEbiography}[{\includegraphics[width=1in,height=1.25in,clip,keepaspectratio]{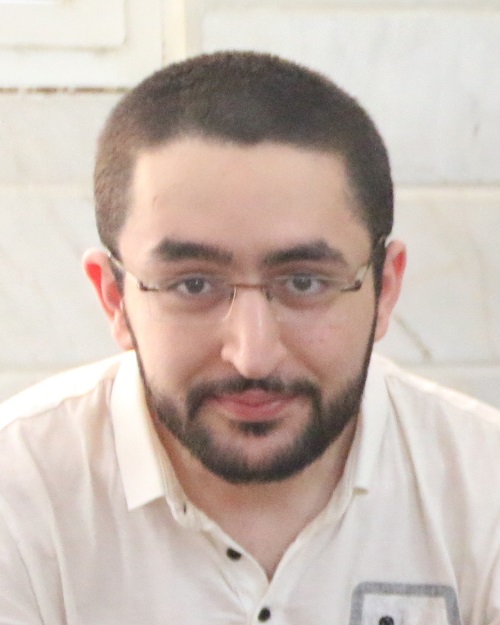}}] {Soheil Shahrouz} received the B.Sc. degree in electrical engineering from Amirkabir University of Technology, Tehran, Iran in 2018. He is currently working towards the M.Sc. degree in electrical engineering at Sharif University of Technology, Tehran, Iran. His research interests include parallel computing and hardware acceleration.
\end{IEEEbiography}

\begin{IEEEbiography}[{\includegraphics[width=1in,height=1.25in,clip,keepaspectratio]{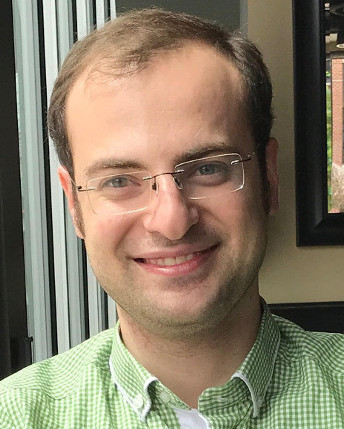}}] {Saber Salehkaleybar} received the B.Sc., M.Sc. and Ph.D. degrees in electrical engineering from Sharif University of Technology, Tehran, Iran, in 2009, 2011, and 2015, respectively. He was a postdoctoral researcher in Coordinated Science Lab. (CSL) at University of Illinois, Urbana-Champaign in 2016-2017. He is currently an assistant professor of electrical engineering at Sharif University of Technology, Tehran, Iran. His research interests include distributed systems, machine learning, and causal inference.
\end{IEEEbiography}

\begin{IEEEbiography}[{\includegraphics[width=1in,height=1.25in,clip,keepaspectratio]{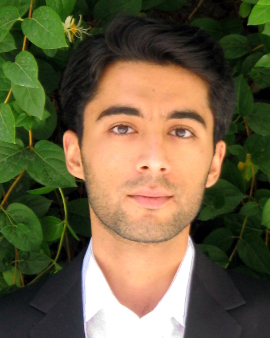}}] {Matin Hashemi} received the B.Sc. degree in electrical engineering from Sharif University of Technology, Tehran, Iran, in 2005, and the M.Sc. and Ph.D. degrees in computer engineering from University of California, Davis, in 2008 and 2011, respectively. He is currently an assistant professor of electrical engineering at Sharif University of Technology, Tehran, Iran. His research interests include algorithm design and hardware acceleration for machine learning and big data applications.
\end{IEEEbiography}





\end{document}